\documentclass[reviewcopy,12pt]{elsart}
\usepackage{amsfonts}
\usepackage{amsmath}
\usepackage{graphicx}
\usepackage{bm}
\usepackage{amssymb}
\usepackage{times}
\usepackage{dcolumn}
\usepackage{psfig}
\usepackage{threeparttable}
\makeatletter

\begin{document}

\begin{frontmatter}

\title{Electronic and magnetic properties of twisted graphene nanoribbon and
M\"{o}bius strips: first-principles calculations}


\author[label1]{Sheng-Ying Yue}
\author[label1,label2]{Qing-Bo Yan}
\author[label1,label3]{Zhen-Gang Zhu}
\ead{zgzhu@ucas.ac.cn}
\author[label1]{Hui-Juan Cui}
\author[label1]{Qing-Rong Zheng}
\ead{qrzheng@ucas.ac.cn}
\author[label1]{Gang Su\corauthref{GSU}}
\corauth[GSU]{Corresponding author. Fax: +86 10 8825 6006.}
\ead{gsu@gucas.ac.cn}
\address[label1]{Theoretical Condensed Matter Physics and Computational Materials Physics Laboratory, School of Physics, University of Chinese Academy of Science, Beijing 100049, China}
\address[label2]{College of Materials Science and Opto-Electronic Technology, University of Chinese Academy of Sciences, Beijing 100049, China}
\address[label3]{School of Electronic, Electrical and Communication Engineering, University of Chinese Academy of Sciences, Beijing 100049, China}

\begin{abstract}
The geometrical, electronic, and magnetic properties of twisted zigzag-edged
graphene nanoribbons (ZGNRs) and novel graphene M\"{o}bius strips (GMS) are
systematically investigated using first-principles density functional
calculations. The structures of ZGNRs and GMS are optimized, and their
stabilities are examined. The molecular energy levels and the spin polarized density of states are calculated. It is found that for
twisted ZGNRs, the atomic bonding energy decreases quadratically with the
increase of the twisted angle, and the HOMO-LUMO gap are varying in a sine-like
behavior with the twisted angle. The calculated spin densities
reveal that the ZGNRs and GMS have antiferromagnetic ground states, which
persist during the twisting. The spin flips on the zigzag edges of GMS are observed at some positions.

\end{abstract}

\end{frontmatter}


\section{Introduction}
Carbon nanostructures such as carbon fullerenes, nanotubes, and graphene have
attracted great interest in basic research and industrial applications. The
techniques for growth and preparation of carbon nanostructures have gained a great
progress, and the scope of feasible applications of
carbon nanostructures has been enlarged dramatically \cite{R.
Saito,P.M. Ajayan}. Since the first successful isolation of graphene, a
single layer of graphite,  the graphene-based materials have attracted
much attention because of an expectation of a new generation of nanodevices%
\cite{R. Saito,P.M. Ajayan,Geim01,Geim02,Geim03,Geim04} such as the graphene field-effect transistor \cite{Max C. Lemme}, the electromechanical resonators from graphene sheets \cite{J. Scott Bunch}, and so on. Especially, to make spintronic nanodevices the materials should bear net spin and magnetism. This stimulates extensive exploration of
magnetism of carbon-based materials. Currently, besides graphene, some
carbon-based materials such as one-dimensional (1D) and two-dimensional (2D) organic polymers with high spin
have been synthesized and explored \cite{Iwamura}. However, it is still
difficult to synthesize organic polymers with macroscopic magnetization.
In fact, these materials exhibit macroscopic diamagnetism because of the
interactions between molecules that may cause the cancellation of magnetism \cite{Iwamura}. Therefore, in order to obtain the magnetic organic materials, it is important to develop relevant techniques for designing and studying the nature of graphene at the atomic level, including the electronic and magnetic
properties for carbon-based molecules and nanostructures.

It is well known that there are two types of edges in a graphene
nanoribbon (GNR)\cite{Geim01}, namely armchair and zigzag edges. Localized
electronic states at the zigzag edges of GNR have been reported theoretically \cite{K.Kobayashi,Klein}, and later confirmed experimentally by scanning
tunneling microscopy and spectroscopy \cite{Y.Kobayashi,Niimi}. Recently, a number of works have predicted the existence of an antiferromagnetic (AFM) ground state that lays up-spin at one edge and down-spin at the other edge
of a zigzag-edged graphene nanoribbon (ZGNR) \cite{Fujita,Nakada,Son,Jiang,DJiang}. The armchair-edged graphene nanoribbon (AGNR) does not have such a magnetic property. Although such AFM ground state has not been confirmed
experimentally for a single graphene nanoribbon, Enoki and coworkers
have observed the localized spins in graphite nanodomains of activated carbon
fibers, and attributed the origin to the zigzag edge \cite{Shibayama,Enoki}.
However, the above-mentioned theoretical and experimental works did not study the
magnetic properties of the twisted ZGNR.

If we connect the two ends of a twisted ZGNR, a carbon M\"{o}bius strip
could be formed. In 1982, the molecules with the shape of a half-twisted M\"{o}bius
strip have been synthesized for the first time \cite{Walba}. In 1998,
Mauksch \textit{et al}. \cite{Mauksch} have presented a computational
reinterpretation of experimental data, showing that $(CH)_{9}^{+}$ could be a
M\"{o}bius aromatic cyclic annulene with 4n $\pi $-electrons. Electronic
properties of ring compounds were discussed theoretically in connection with
M\"{o}bius aromatic properties in some works \cite%
{Santamaria,Rzepa,SRzepa,Kastrup}. Mart\'{\i}n-Santamaria and Rzepa\cite%
{Santamaria} presented an analysis of the main features of M\"{o}bius
annulenes by considering the $\pi $-molecular orbital correlation between the
planar H\"{u}ckel configuration and the twisted $C_{2}$ symmetric M\"{o}bius
system. The synthesis of a stable M\"{o}bius aromatic hydrocarbon has been
obtained for the first time in 2003 by Ajami \textit{et al}. \cite{Ajami}
who combined a normal aromatic structure (such as benzene, with trigonal
planar $sp^{2}$-hybridized atoms) and a belt-like aromatic structure (such as
the surface of a carbon nanotube, with pyramidalized $sp^{2}$ atoms),
creating a M\"{o}bius compound stabilized by an extended $\pi $-system. In
2008, Caetano {\it et al.} have investigated theoretically the structural stability,
geometrical optimization and electronic properties of twisted graphene
nanoribbon and M\"{o}bius-like graphene rings \cite{Caetano}, but magnetic
edge states have not been touched.

In this paper, the geometric, electronic and magnetic properties of twisted
ZGNRs and M\"{o}bius strips will be systematically investigated by means of the first
principles calculations. For the ZGNRs, various sizes and twisted angles will be
considered, and the main focus will be given to the effect of different twisted
angles on electronic and magnetic properties of ZGNRs. In addition, a
series of M\"{o}bius-like strips derived from ZGNRs will be generated, and their
electronic and magnetic properties will be also studied.

\section{Computational Details}

All calculations have been performed using the Vienna Ab-initio Simulation
Package (VASP) \cite{Kresse G01,Kresse G02} code, implementing the
spin-polarized density functional theory (DFT) and the GGA of Perdew and
Wang \cite{Perdew J. P.} known as PW91. We have used projector augmented
wave (PAW) potentials \cite{Kresse G,Blochl} to describe the core ($1s^{2}$)
electrons, with the $2s^{2}$ and $2p^{2}$ electrons of carbon considered as
valence electrons. A kinetic energy cutoff $400$ eV is taken. Only $\Gamma $
Point is used for $k$- sampling. The force tolerance is set at $0.025$ eV/{%
\AA }. The spin charge densities are calculated by taking the difference of
the spin up and spin down charge densities. The supercells are created by
including a vacuum layer with thickness larger than $10$ {\AA }. In
order to show the edge magnetic properties clearly, the zigzag edges of the
ZGNRs and M\"{o}bius-like strips are not hydrogenated.

\section{Results and Discussion}

\subsection{Structures and Stability}

\begin{figure}[tbp]
\includegraphics[width=0.8\linewidth,clip]{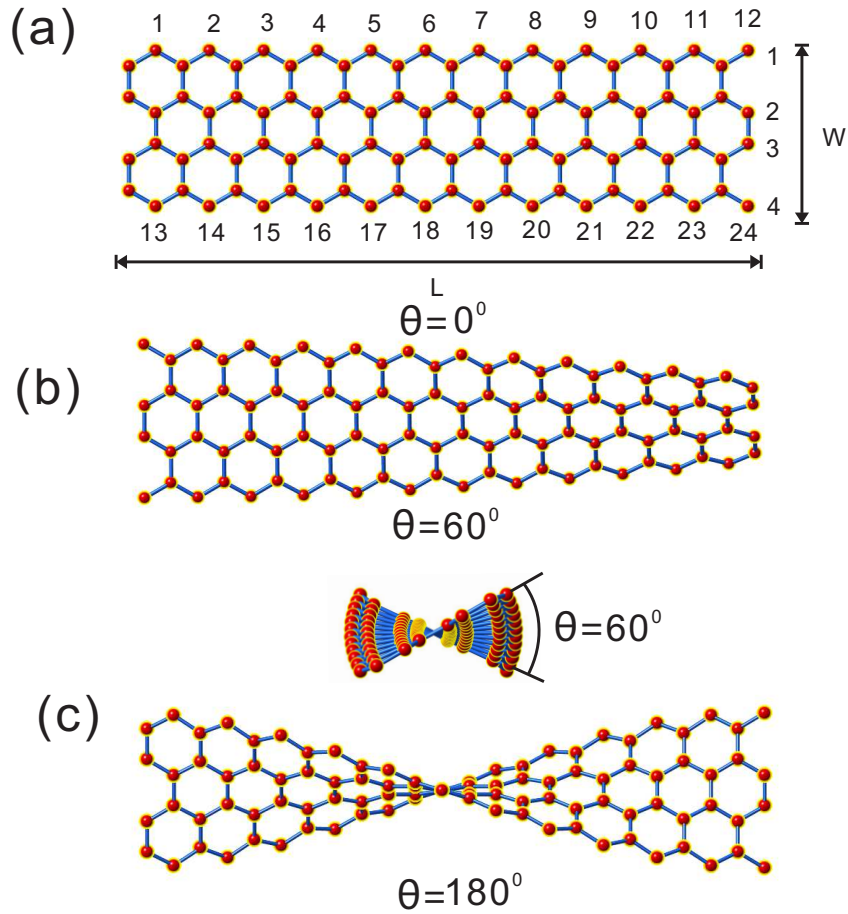}
\caption{The structures of untwisted and twisted zigzag-edged graphene nanoribbons (ZGNRs), where L and W are the length and width of ZGNRs, respectively, and $\protect\theta$ is the twisted angle. (a) $\protect\theta=0^{\circ}$; (b) $\protect\theta=60^{\circ}$; (c) $\protect\theta=180^{\circ}$.}
\label{1}
\end{figure}

\begin{figure}[tbp]
\includegraphics[width=1.0\linewidth,clip]{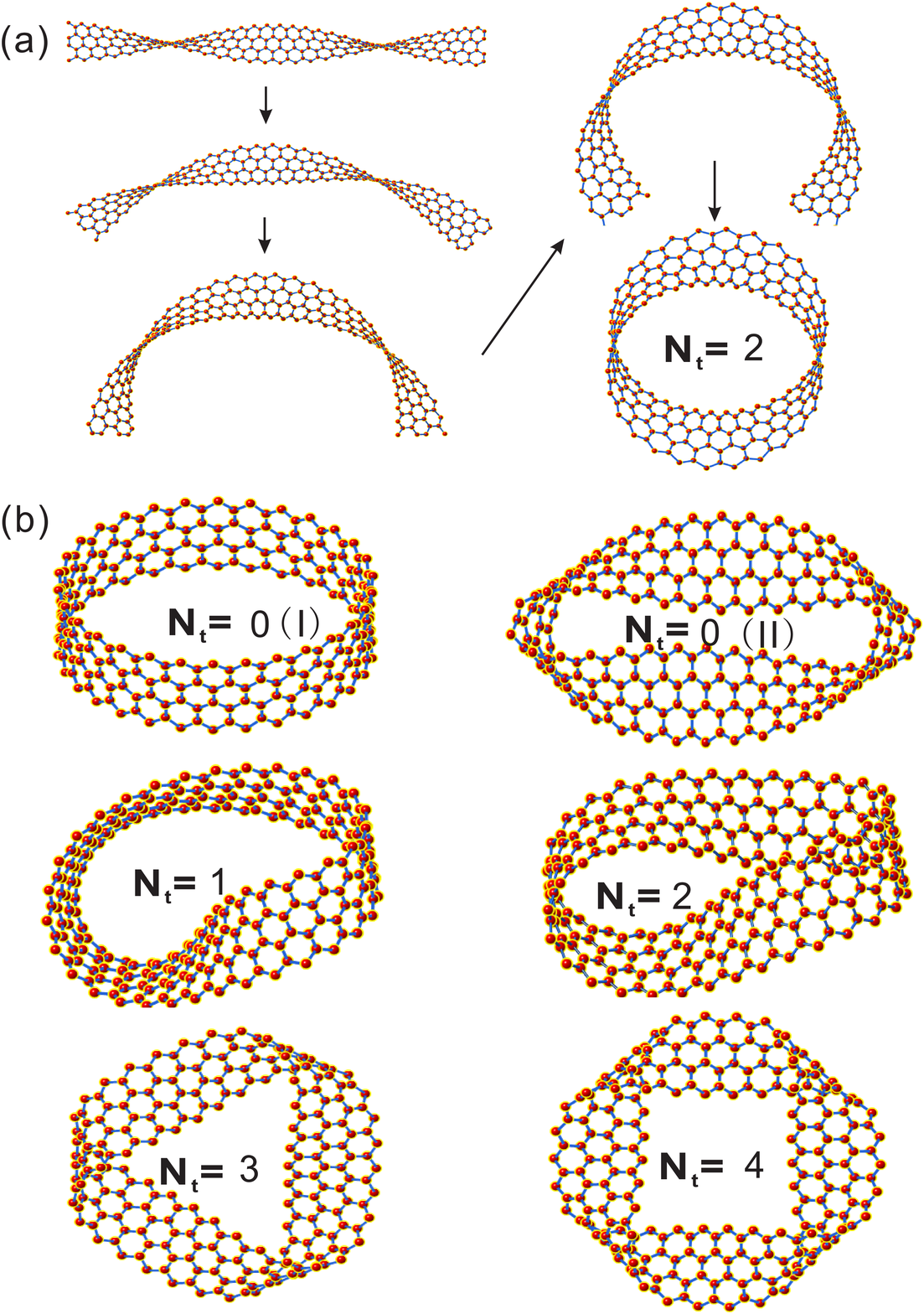}
\caption{(a) An illustration how to generate $N_{t}=2$ M\"{o}bius
strip by using a twisted ZGNR with L=30, W=4, and N=240. (b) Several M\"{o}bius
strips generated by twisted ZGNRs with L=30, W=4, and N=240, where $N_{t}$ is the twisted
times. When $N_{t}=0$, there exist two stable nano-ring strip structures. All
structures presented here were optimized with the DFT and the MD simulated annealing and molecular
mechanics.}
\label{2}
\end{figure}

First let us cut graphene to generate ZGNRs with specific
sizes, and then twist them with different angles. Fig. \ref{1}(a) is the
structure of untwisted ZGNR with the number of carbon atoms $N=96$, the length $L=12$
and the width $W=4$. Figs. \ref{1}(b) and \ref{1}(c) give the twisted ZGNRs
with twisted angles $\theta =60^{\circ }$ and $180^{\circ }$, respectively.
To investigate systematically the twisted ZGNRs, various twisted structures
are considered, ranging from $L=6$ to $12$, $\theta =0^{\circ }$
to $180^{\circ }$, and $N=48$ to $96$, while $W$ is kept at 4. Besides, the
structures with $W=8$, $N=96$, $\theta =0^{\circ }$ to $180^{\circ }$ are also
included to inspect the effect of different $W$ on the physical properties of ZGNRs (Supplemental
Materials). All of the geometrical structures of twisted ZGNRs are relaxed,
except for the atoms on the armchair edges (the two ends of ZGNRs), as
they should be fixed in experiments.

If the two armchair ends of the twisted ZGNRs are connected together using carbon-carbon
(C-C) bonds, M\"{o}bius-like strips of graphene could be obtained, as shown in
Fig. \ref{2}(a). Obviously, different twisted angles of generated structures
bear different geometric topologies, which could be expressed by the
twisted times $N_{t}$, and the twisted angle $\theta =N_{t}{\times }%
180^{\circ }$. If $N_{t}=0$, a short carbon nanotube could be formed, as indicated
in Fig. \ref{2}(b). If $N_{t}=1$, the conventional M\"{o}bius strip can be obtained. Furthermore, Fig. \ref{2}(b) also presents more M\"{o}%
bius-like strips with higher twisted times $N_{t}=2$, $3$, and $4$. In
particular, the structures of $N_{t}=0$ to $4$, $L=30$ to $40$, $W=4$, $N=240$
to $320$ can be generated (only the structure with $N=240$ is shown in Fig. ~\ref{2}(b)). To
check the dynamical stability and to relax the geometry, molecular dynamics
(MD) calculations have been performed to simulate the annealing process from
$1000$ K to $300$ K. It turns out that all generated structures are
kept, and only small deformations occur. Interestingly, a new metastable
structure with $N_{t}=0$ is also uncovered in the simulated annealing process
(the second one of Fig. \ref{2}(b)).

For the above twisted ZGNRs and M\"{o}bius-like strips, the structures are
optimized again, and their relative energies are obtained. To compare the relative
energies of structures with different number of carbon atoms, the atomic bonding energy
$E_{b}$ is a proper quantity, which is defined as
\begin{equation}
-E_{b}=\frac{E_{\text{tot}}}{N}-E_{c},  \label{eb}
\end{equation}%
where $E_{\text{tot}}$ is the total free energy of the system, $N$ is the
total number of atoms, and $E_{c}$ is the energy of isolated carbon atom. The
lower the $-E_{b}$ (or the larger the $E_{b}$), the more stable the structure.

\begin{figure}[tbp]
\includegraphics[width=0.85\linewidth,clip]{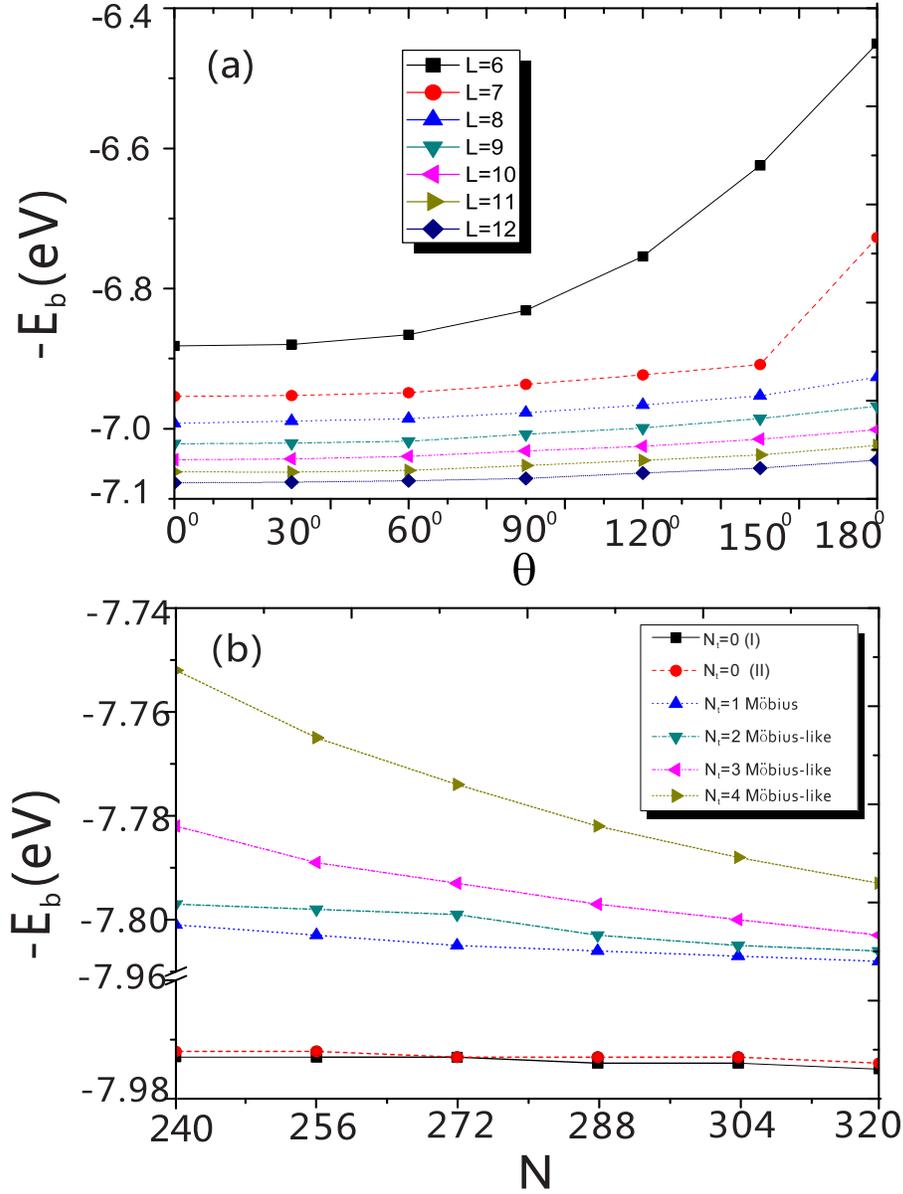}
\caption{The negative atomic bonding energy $E_{b}$ of the twisted ZGNRs versus (a) the twisted angle $\protect\theta$ for different length $L$ and (b) the number $N$ of carbon atoms for different twisted times $N_t$, where the width of the ZGNRs $W=4$.}
\label{3}
\end{figure}

The atomic bonding energies of the untwisted and twisted ZGNRs are given in Table. \ref{tab1}. Fig. \ref{3}(a) shows $E_{b}$ of several untwisted and twisted ZGNRs with the length from $L=6$ to $12$
and the twisted angle from $\theta =0^{\circ }$ to $180^{\circ }$. It is obvious that
$-E_{b}$ of the twisted ZGNRs are increased with the increase of the twisted
angle $\theta $ in a quadratic way, which could be easily understood that
the twisting enhances the strain of ZGNRs, and thereby raises the elastic energy.
Moreover, $-E_{b}$ of structures with shorter $L$ are generally larger than
that of longer structures at the same twisted angle, and $-E_{b}$ with
shorter $L$ grows more steeper than those with larger $L$, which is understandable
as the shorter ZGNRs present a larger strain at the same twisted angle.
It is clear that the longer ZGNRs are more stable, while the twisting
reduces the energy stability of ZGNRs.

The atomic bonding energy $E_{b}$ of M\"{o}bius-like strips with twisted
times from $N_{t}=0$ to $4$ and the number of atoms from $N=240$ to $360$ are listed in
Table \ref{tab2}. Generally, $-E_{b}$ with the
same $N_{t}$ are decreasing with the increase of the number of atoms $N$, and $%
-E_{b}$ with larger $N_{t}$ grows steeper than those with smaller $N_{t}$
[Fig. \ref{3}(b)], which could be attributed to the fact that a larger twisted times brings a higher strain
at the same strip size. For $N_{t}=0$, the bonding energies $E_{b}$ are much
larger than those of the cases with $N_{t}>0$.

It can be seen from Fig. \ref{3}(b) that $E_{b}$ with $N_{t}=0$ are
highly competitive, which shows a similar trend for the new stable
structure (as labeled in Fig. \ref{2} with $N_{t}=0$ (II)) and the nanotube (as in Fig. \ref{2} with $N_{t}=0$ (I)). For M\"{o}bius-like strips with $N_{t}>0$, a
larger $N_{t}$ leads to a less stable structure (with a larger $-E_{b}$). For a fixed $%
N_{t}$, we observe that the structures with smaller N are more unstable because a smaller nano-ring bears a larger strain from the twisting process. This is also manifested in $E_{b}$ that tends to a common value for bigger M%
\"{o}bius-like strips since the strain effect becomes smaller in the larger
system. We should point out that all the structures studied here are optimized by the first-principles calculations.

\begin{table*}[tbp]
\caption{$E_{b}$ $(eV)$ of untwisted and twisted ZGNRs.}
\label{tab1}\centering
\begin{tabular*}{17cm}{@{\extracolsep{\fill}}lcccccccccc}
\hline\hline
Twisted angle & N=48 & N=56 & N=64 & N=72 & N=80 & N=88 & N=96 &  &  &  \\
\hline
$\theta=0^{\circ}$ & -6.882 & -6.954 & -6.992 & -7.021 & -7.044 & -7.062 &
-7.077 &  &  &  \\
$\theta=30^{\circ}$ & -6.880 & -6.953 & -6.989 & -7.020 & -7.043 & -7.062 &
-7.076 &  &  &  \\
$\theta=60^{\circ}$ & -6.866 & -6.948 & -6.986 & -7.018 & -7.040 & -7.059 &
-7.074 &  &  &  \\
$\theta=90^{\circ}$ & -6.831 & -6.937 & -6.977 & -7.008 & -7.032 & -7.052 &
-7.071 &  &  &  \\
$\theta=120^{\circ}$ & -6.754 & -6.923 & -6.966 & -6.999 & -7.025 & -7.045 &
-7.063 &  &  &  \\
$\theta=150^{\circ}$ & -6.624 & -6.908 & -6.953 & -6.985 & -7.015 & -7.037 &
-7.056 &  &  &  \\
$\theta=180^{\circ}$ & -6.450 & -6.727 & -6.926 & -6.968 & -7.001 & -7.024 &
-7.045 &  &  &  \\ \hline\hline
\end{tabular*}%
\end{table*}

\begin{table*}[tbp]
\caption{$E_{b}$ $(eV)$ of graphene nano-rings with the atomic number $N=240-320$ and the
twisted times $N_{t}=0-4$.}
\label{tab2}\centering
\begin{tabular*}{17cm}{@{\extracolsep{\fill}}lcccccccccc}
\hline\hline
N & Nanotube & Novel Ring & M\"{o}bius $N_{t}=1$ & $N_{t}=2$ & $N_{t}=3$ & $%
N_{t}=4$ &  &  &  &  \\ \hline
240 & -7.973 & -7.972 & -7.800 & -7.797 & -7.782 & -7.752 &  &  &  &  \\
256 & -7.973 & -7.952 & -7.792 & -7.798 & -7.789 & -7.766 &  &  &  &  \\
272 & -7.973 & -7.973 & -7.804 & -7.799 & -7.793 & -7.774 &  &  &  &  \\
288 & -7.974 & -7.963 & -7.804 & -7.803 & -7.797 & -7.782 &  &  &  &  \\
304 & -7.974 & -7.963 & -7.807 & -7.798 & -7.800 & -7.788 &  &  &  &  \\
320 & -7.975 & -7.974 & -7.808 & -7.801 & -7.803 & -7.793 &  &  &  &  \\
\hline\hline
\end{tabular*}%
\end{table*}

\subsection{Electronic Structures of Twisted ZGNRs}

\begin{figure}[tbp]
\includegraphics[width=0.85\linewidth,clip]{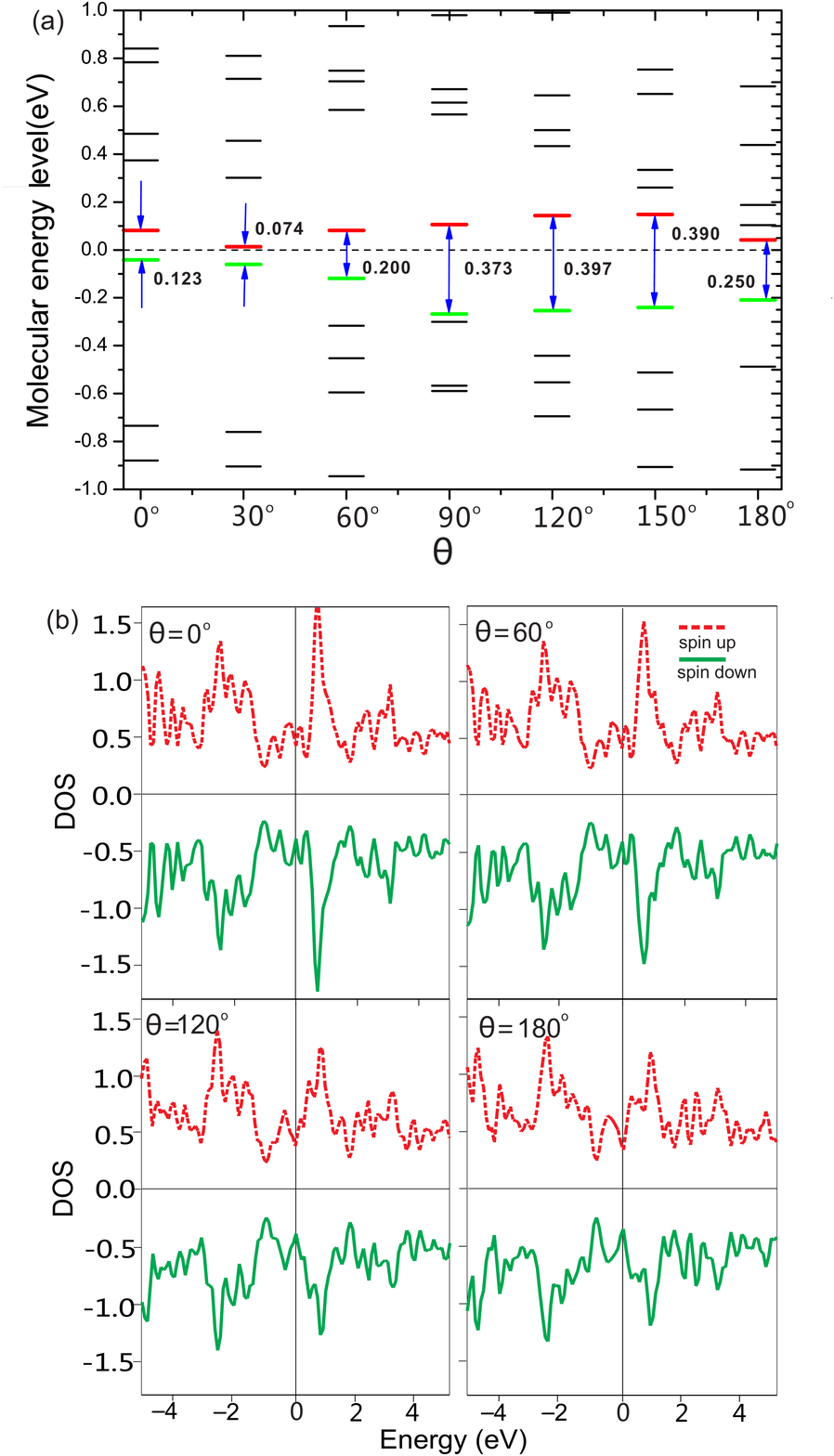}
\caption{(a) The molecular energy levels of the ZGNRs versus the twisted angle
$\protect\theta$. (b) The spin polarized DOS of the ZGNRs for different twisted angle $\protect\theta$%
, where the red dot lines represent the DOS of electrons with spin up and the green solid lines for the electrons with
spin down. The Fermi level is set at the zero energy point, and L=12, W=4 and N=96 are taken.}
\label{4}
\end{figure}

The electronic structures of twisted ZGNRs are investigated. The
molecular energy levels and the density of states (DOS) with different
twisted angle $\theta $ are shown in Fig. \ref{4}. The energy gap is also
obtained, which is defined by an energy difference between the lowest unoccupied
molecular orbit (LUMO) and the highest occupied molecular orbit (HOMO) of
the graphene ZGNRs. The smallest gap $0.074$ eV occurs at the twisted angle $%
\theta =30^{\circ }$, while the largest energy gap is $0.397$ eV, which appears at the twisted angle $\theta =120^{\circ }$ [Fig. \ref{4}(a)]. Generally, the energy
gap becomes larger but then decreases as the twisted angle grows, showing a $%
sine$-like behavior. The fitting equation has the form of
\begin{equation}
E_{\text{gap}}=E_{\text{gap},0}+A\sin \left[ \frac{\pi (\theta -\theta _{c})%
}{w}\right],   \label{egap}
\end{equation}
where $E_{\text{gap}}$ is the HOMO-LUMO energy gap, and the fitting parameters are $E_{%
\text{gap},0}=0.258$ eV, $\theta _{c}=71.936^{\circ }$, $A=0.171$ eV and $%
w=105.676^{\circ }$. The results of the HOMO-LUMO gap reveal
that the twisted ZGNRs could be tunable organic semiconductor materials.

In Fig. \ref{4}(b), the DOS of electrons with spin up (red dashed lines) and spin down (green solid lines) are shown for the twisted angles $\theta=0^{\circ }$, $60^{\circ }$, $120^{\circ }$ and $180^{\circ }$, respectively. It is seen that the magnitude of DOS for up and down spins coincide with each other for different twisted angles, implying that ZGNRs ground states are antiferromagnetic or nonmagnetic. The DOS profile reveals that the small peaks in the spin-up subbands just above
and below the Fermi level move away from the Fermi level upon applying the twist.
This could be caused by the charge transfer from the local edge to the middle of the graphene nanoribbon. The reason is that the strain induced by the twisting changes the bond length and the bond angles. In accordance with this change, the charge may be transferred from one carbon atom to others since the originally orthogonal orbitals are hardly to maintain themselves in the same orientation after relaxation. As a result, the $\pi _{z}$ orbitals at the ribbon edge are no longer orthogonal to the $\sigma $ orbitals, and are not exactly parallel to each other among themselves with increasing the twisted angle. When this happens, the magnitudes of the two small peaks just above and just below the Fermi level are increased and the peaks are sharpened.\\

\subsection{Edge Magnetism of Twisted ZGNRs and Graphene M\"{o}bius-like strips}

\begin{figure}[tbp]
\includegraphics[width=0.9\linewidth,clip]{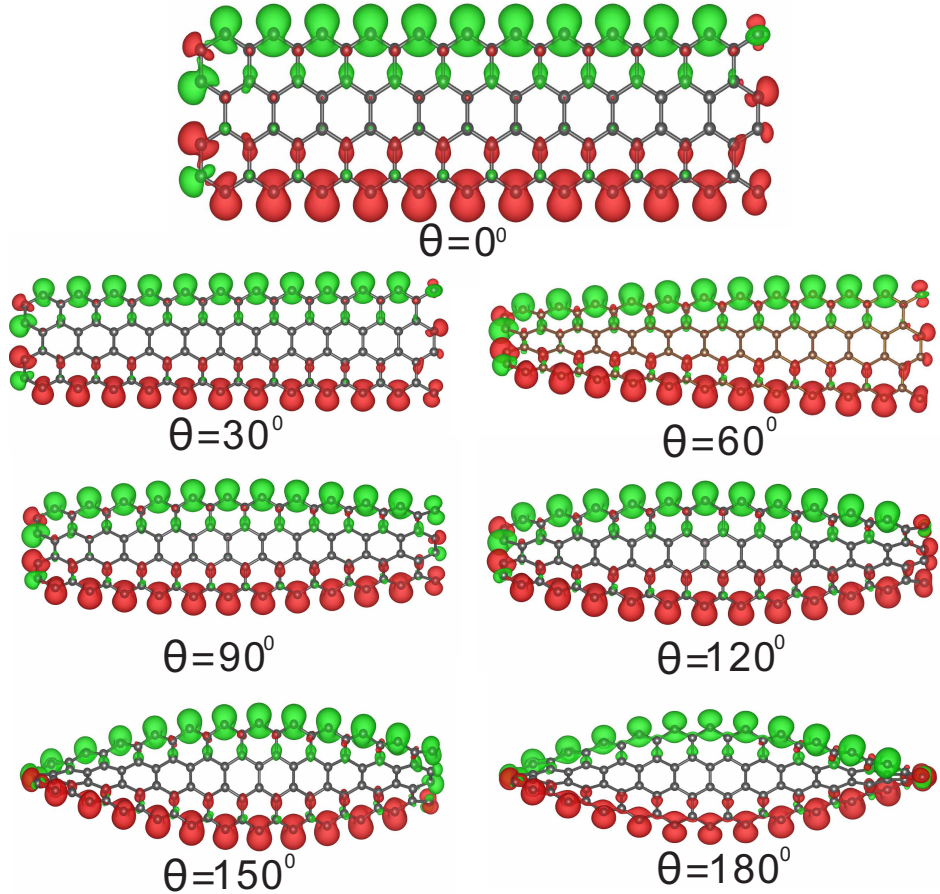}
\caption{Spin densities of twisted ZGNRs with L=12, W=4, and N=96 for different twisted angles, where green
isosurface represents the charge density of electrons with spin up, and red isosurface represents the charge density of electrons with spin down.}
\label{5}
\end{figure}

\begin{figure}[tbp]
\includegraphics[width=0.9\linewidth,clip]{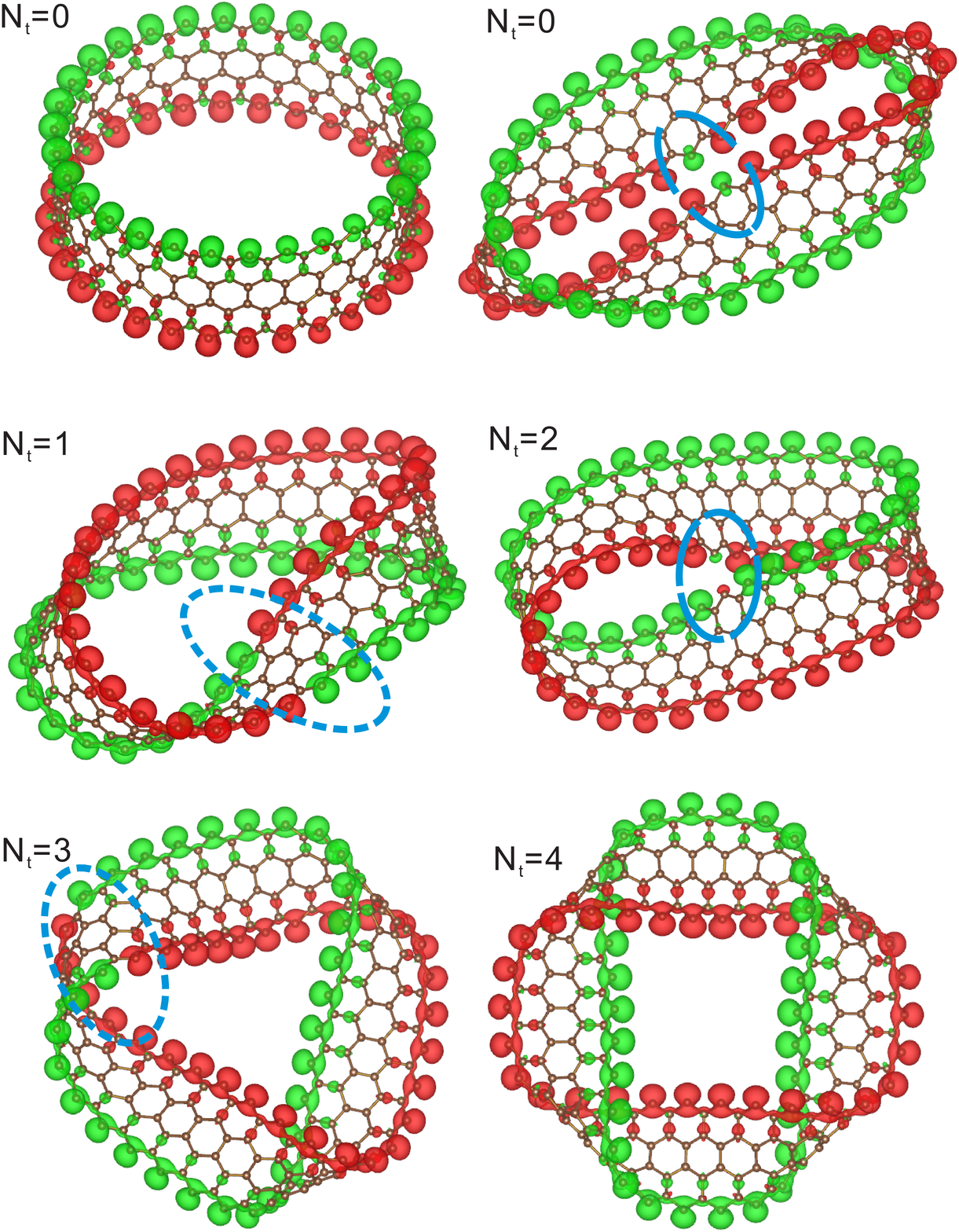}
\caption{Spin densities of zigzag-edged graphene M\"{o}bius-like nano strips (with L=30, W=4, N=240) for different geometric topologies, where green isosurface represents the charge density of electrons with spin up, and red isosurface represents the charge density of electrons with spin down.}
\label{6}
\end{figure}

The charge transfer may not only lead to charge redistribution but also
redistribution of the local magnetic moments in the real space. The latter could
manifest itself in a variation of the original antiferromagnetic edge
magnetism of ZGNRs. In this subsection, the effect of twisting on
spin density distribution ($\rho_{\uparrow} -\rho_{\downarrow} $) of
twisted ZGNRs and M\"{o}bius-like strips are explored, where $\rho_{\uparrow}$ and $\rho_{\downarrow}$ are the charge densities of electrons with spin up and spin down, respectively. The results are presented in Figs. \ref{5} and \ref{6}. It is known that each carbon atom at zigzag edge of graphene possesses
four valence electrons, two of which participate in the $\sigma$ covalent
bonds forming the honeycomb structure of graphene, while the other two are
unpassivated electrons, one of which forms the dangling $sp^{2}$ bond, and another is in
the dangling $p_{z}$ orbital. Without twisting, this $p_{z}$ orbital is
orthogonal to the surface orbitals and localized at the edge. The dangling $%
sp^{2}$ bond is probably spin-polarized slightly \cite{Lehtinen}.

Fig. \ref{5} demonstrates the spin densities of ZGNRs at different twisted angles. We can see that the ZGNRs always take the AFM edge ground states that are manifested by the opposite net spin densities at zigzag edges during twisting whenever the twisted angle is small or large, where the spin density (local magnetic moment) is obtained by integrating \textit{spd}-projected DOS up to the Fermi level. This result is also consistent well with the spin DOS in Fig. \ref{4}(b), showing that twisting does not have a qualitative effect on the AFM ground states of ZGNRs.

There are a few previous studies on spin polarized properties \cite{Jiang}
and structural stability \cite{Caetano} of graphene M\"{o}bius-like strips,
but the spin-resolved edge states of M\"{o}bius-like nano strips with higher
twisted times have not yet been reported. Therefore, it is interesting to pay attention to the nontrivial question whether twisting can affect the magnetic property of the zigzag-edged graphene M\"{o}bius-like strips, particularly when M\"{o}bius-like strips with odd twisted times $N_{t}$ only have a single edge in geometric sense. The calculated results are given in Fig. \ref{6}. When $N_{t}$ is odd, the M\"{o}bius-like strips also
keep the AFM ground states, but the spin flips can happen at some contrapositions of the two nano-ribbon edges, as indicated by blue-dashed circles in Fig. \ref{6}.
When  $N_{t}$ is even, M\"{o}bius-like strips still retain the AFM ground states, and in some positions spin flips can also appear. However, the spin flip in this case happens not at the contrapositions of the
edges but at those positions where two carbon atoms are the closest to each other in real space after relaxation, which are marked by blue-long-dashed circles for $N_{t}=0,2$ in Fig. \ref{6}. This phenomenon may be caused by electronic interactions of carbon atoms.

\section{Summary}

In this work we have investigated the structural, electronic and magnetic properties of
the twisted zigzag-edge graphene nanoribbons (ZGNRs) and M\"{o}bius-like
strips built on ZGNRs by using the first-principles DFT calculations. The stabilities of
those novel structures are examined. The molecular energy levels and the spin polarized density of states are obtained. It is found that the atomic bonding energy of
the twisted ZGNRs decreases quadratically with the increase of the twisted
angle. The molecular HOMO-LUMO energy gap is disclosed to vary in a sine-like behavior with
the twisted angle. The charge transfer is observed, which is due to the twisting on nanoribbons. The twisted ZGNRs could
be taken as tunable organic semiconductor materials. The
energy of graphene M\"{o}bius-like strips is uncovered to increase with the twisted
times $N_{t}$ and the effect induced by the twisting process is degraded with
increasing of the number of atoms. By using the molecular dynamics
simulating annealing process and DFT optimization we also find a new stable
structure of graphene short-nanotube. The spin-polarized DFT calculations
show that the AFM ground states of the twisted ZGNRs and graphene M\"{o}bius-like strips are persistent during twisting.
In addition, the spin states on the zigzag edges of M\"{o}%
bius-like strips are observed to flip at some positions. For the ring strip with odd $N_{t}$, the flip
occurs at the contrapositions of the two zigzag edges. For the strips with $%
N_{t}=0$ and $2$, the flip takes place at those where two carbon atoms are
placed at the closest positions after structure optimization. Our present study might be helpful to understand further the properties of graphene at atomic scales and is useful for future designs of spintronic nanodevices.
\bigskip

\section*{Acknowledgements}

All calculations are
completed on the supercomputer NOVASCALE7000 in Computer Network Information
Center (Supercomputing center) of Chinese Academy of Sciences (CAS) and MagicCube
(DAWN5000A) in Shanghai Supercomputer Center. This work is
supported in part by the NSFC (Grants No. 90922033, No. 10934008, No.
10974253, No. 11004239), the MOST of China (Grant No. 2012CB932901, No. 2013CB933401), and the CAS.

\bigskip
\section*{Appendix A. Supplementary data}
Supplementary data associated with this article can be found, in the
online version, at doi:******.

\end{document}